# Antiferroelectric and Magnetodielectric Coupling Response of $La_{0.2}Sr_{0.7}Fe_{12}O_{19}$ Ceramics

Yao Huang, Guolong Tan[*]

State Key Laboratory of Advanced Technology for Materials Synthesis and Processing,

Wuhan University of Technology, Wuhan 430070, China

## Abstract

Multiferroics is a class of functional materials that simultaneously exhibit ferroelectricity and ferromagnetism in a single structure. We report here the integration of anti-ferroelectricity and ferromagnetism in a new M-type hexaferrite compound, in which 0.3 Sr ions was substituted by 0.2 $La^{3+}$ ions in $SrFe_{12}O_{19}$ so as to keep the charge balance. This doped compound is expressed as a molecular formula of $La_{0.2}Sr_{0.7}Fe_{12}O_{19}$, which was confirmed by XRD to have the same structure as $SrFe_{12}O_{19}$ with a lattice contraction of 0.59%. Surprisingly, the doping effect turns $La_{0.2}Sr_{0.7}Fe_{12}O_{19}$ from ferroelectric to antiferroelectric phase, which displays double hysteresis loops with a maximum polarization of 154 $\mu C/cm^2$ and a remnant one of 38 $\mu C/cm^2$. The dielectric constant $\varepsilon^{'}$ is also greatly improved from 1462 to 1451778 at 10.2Hz by this substitution. This compound simultaneously demonstrates strong ferromagnetism, the remnant magnetic moment and coercive field are 52 emu/g and 5876 Oe, respectively. In addition, $La_{0.2}Sr_{0.7}Fe_{12}O_{19}$ also exhibits strong magnetodielectric (MD) response, in which a applying magnetic field B lifts the whole $\varepsilon^{'}$-f spectra up and right shifts $\varepsilon^{''}$-f spectra $(\omega_m)$ to higher frequency side. Both real and imaginary parts of the dielectric constant varies with frequency and B field and obey Debye relaxation model. The maximum MD increment in $\varepsilon^{'}$ reaches as high as 540830 upon a B field of 926mT. The capacitance of the $La_{0.2}Sr_{0.7}Fe_{12}O_{19}$ ceramics could be much more enhanced by applying a magnetic field B, which could induce an additional polarization P(H) upon the conventional P(E) by the cycloid conical spin in the intermediate phases. These results suggest that $La_{0.2}Sr_{0.7}Fe_{12}O_{19}$ exhibits a strong interplay between magnetic ordering and ferroelectricity, which makes it a good magnetoelectric coupling candidate.

## 1. Introduction

[*] Corresponding author: gltan@whut.edu.cn, Tel: 0086-27-87651837-8405, Fax: 0086-27-87879468

Ferroelectrics are materials that display an electric polarization in zero applied electric field in which the direction of the electric polarization is reversed by the electric field. Ferroelectric order results from the formation of an electric moment within the unit cell of the crystal and can only exist in a material with broken inversion symmetry. Ferromagnets are magnetically ordered materials in which all the magnetic moments are aligned in the same direction in zero applied magnetic field and in which time-reversal symmetry is broken. Multiferroics was originally coined for materials in which ferroelectrics and ferromagnets coexist in the one single phase. Multiferroic (MF) materials have attracted certain attentions in the past decade due to their interesting physics and multifunctional performances such as spintronic devices, nonvolatile memory devices, solid-state transformers as well as magnetoelectric (ME) effects where the electric polarization ($P$) can be tuned by magnetic field ($H$) or magnetization ($M$) can be tuned by electric field ($E$).[1, 2, 3, 4, 5, 6, 7] For technological point of view, joint occurrence of large ferroelectric and strong ferromagnetic polarizations as well as strong magnetoelectric coupling in single MF material are highly desired for practical application. However, these two merits are rarely found to be compatible in the single-phase MF materials[7], since ferroelectricity and magnetism have contradicting requirements. While magnetism requires d electrons, ferroelectric distortion is suppressed by d electrons and requires non central symmetric space group. Such incompatibility provides a strict limitation on discovery of new multiferroic materials and would somehow prevent their practical applications.

Despite the limitations mentioned above, several multiferroics have been discovered wherein ferroelectricity is induced by ingeneous routes.[2, 3, 4] Since the discovery of multiferroic behaviors in perovskite $BiFeO_3$ and $TbMnO_3$[8, 9], a large number of multiferroic materials with different physical mechanisms have been found in the past decade,[5, 10, 11] including $LuFe_2O_4$,[12, 13] $LaMn_3Cr_4O_{12}$[6] and $BiMn_3Cr_4O_{12}$,[7] magnetically induced ferroelectric $TbMnO_3$,[14] $GdFeO_3$,[15] $SmFeO_3$,[16] $LnMn_2O_5$[17] as well as Y-type and Z-type hexaferrites[1, 10, 18, 19,] which are termed as type-II multiferroics. In $YMnO_3$ and related manganites, ferroelectricity results from polyhedral tilting.[20] Lone pair (6s) electrons play a role in imparting ferroelectricity in bismuth-containing transition-metal oxides.[21] Frustrated magnetism and spiral magnetic ordering induces ferroelectricity in manganese oxides of the type $LnMn_2O_5$ (Ln=rare earth)[14] and $TbMnO_3$.[17] Magnetic field induced ferroelectricity in Y-type, M-type and Z-type hexaferrites has become an intense area of investigation in the past few years[22, 23, 24, 25, 26, 27, 28].

Recently, prominent multiferroic performance were discovered in M-type Hexaferrites ($PbFe_{12}O_{19}$[29, 30], $SrFe_{12}O_{19}$[31, 32]) in which large ferroelectricity and strong magnetism coexisted at room temperature. M-type hexaferrites is a traditional permanent magnet with high saturation magnetization and large magnetic coercivity, and has been widely used in recording media, permanent magnets, and components in microwave and high-frequency devices.[33,34] The ferroelectricity of M-type hexaferrites has been verified by classic polarization hysteresis loops with good saturation, huge abnormal change in dielectric constants near Curie temperature for ferroelectric phase change and two nonlinear I-V peaks for switching of polarization field or domains.[29,31] Magnetoelectric coupling in form of giant magnetocapacity effect has also been discovered in a modified M-type hexaferrite of $La_{0.2}Pb_{0.7}Fe_{12}O_{19}$.[32] The multiferroics of M-type hexaferrites has also been observed by many research groups.[10, 35, 36, 37, 38, 39] This is a type of practicable multiferroic compound which simultaneously breaks the central inversion symmetry and time-reversal symmetry, i.e. unpaired electrons in 4d orbits of $Fe^{3+}$ provides spin for magnetic ordering and the distorted $FeO_6$ octahedron provides electric displacement for ferroelectric polarization. The spin-orbit coupling or Dzyaloshinskii–Moriya (DM) interaction produces magnetoelectric coupling response. Therefore ferroelectricity, ferromagnetism and magnetoelectric coupling could jointly coexist in one structure of M-type hexaferrite at ambient condition.

In this work, ferroelectric and magnetic features of $La_{0.2}Sr_{0.7}Fe_{12}O_{19}$ has been investigated. The magnetodielectric coupling response of this multiferroic compound has also been revealed by frequency and magnetic field dependent dielectric spectra. The mechanism for variation of frequency-dependent dielectric constants upon a applying magnetic field will be discussed in detail.

## 2. Experimental Procedure

Firstly, $La_{0.2}Sr_{0.7}Fe_{12}O_{19}$ fine powders were prepared by a polymer precursor procedure. Strontium acetate ($Sr(CH_3COO)_2 \cdot 3H_2O$), lanthanum acetate ($La(CH_3COO)_2 \cdot 3H_2O$) and ferric acetylacetonate ($C_{15}H_{21}FeO_6$) were used as starting compounds, which were purchased from Alfa Aesar. (0.2627g) strontium acetate and 0.1252g lanthanum acetate ($La(CH_3COO)_2 \cdot 3H_2O$) were dissolved in 15 ml glycerin separately to form two types of clear precursor solutions, which were distilled in a rotary evaporator at 120℃ for 1h and then stored in two glass bottles respectively. The atomic ration of strontium-lanthanum mixture (La+Sr) to iron (Fe) was kept at the range of 1:9.5~1:10, so as to compensate the loss of (Sr or La) evaporation during heat treatment process. Meanwhile, (5.5126g) ferric acetylacetonate was dissolved in a mixture solution of 50mL anhydrous ethanol and 70 mL acetone in a 250mL three-neck flask inside a glove box, in this way moisture induced hydrolysis of ferric acetylacetonate could be avoided.

Stoichiometric strontium and lanthanum precursor solutions were moved into the glove box and then poured into the above ferric precursor solution to form a transparent mixture solution, which was stirred at 70℃ for more than 8h. Afterwards, 45mL ammonia solution with 1g polyethylene glycol was added into the above mixture solution, which was continuously stirred again at 70℃ for 24h to form a dispersion solution. Finally, the dispersion solution was moved out of the glove-box and centrifuged at a speed of 12000rpm for 15min to remove the water and organic molecules. The precipitant was thus separated out and then calcined at 450℃ in a furnace for 1h. The powder was further calcined again at 800℃ for another 1 hour to fully remove the organic molecules. In this way, fine $La_{0.2}Sr_{0.7}Fe_{12}O_{19}$ powders were prepared. After grinding for 1 hour, 0.050g $La_{0.2}Sr_{0.7}Fe_{12}O_{19}$ powders were pressed inside a steel module at a pressure of 20 MPa into a $\phi6\times0.3$ mm pellet sample, which was sintered at 1100℃ for 1h into a ceramic specimen. The ceramic pellet was subsequently annealed at 800℃ in pure $O_2$ atmosphere for 3hs, then turned upside down for another 3hs oxygen heat treatment at the same temperature. After cooling down to room temperature, the $La_{0.2}Sr_{0.7}Fe_{12}O_{19}$ pellet was annealed at 700℃ for 3hs in $O_2$ atmosphere one more time so as to fully remove the oxygen vacancies and transform $Fe^{2+}$ into $Fe^{3+}$. Phase identification of the as prepared $La_{0.2}Sr_{0.7}Fe_{12}O_{19}$ powders were performed by X-ray powder diffraction (XRD) using Cu-$K_\alpha$ radiation. For ferroelectric measurement, both surface sides of the ceramic pellet were coated with silver paste as the electrodes, which was heat treated at 820℃ for 15 min. Then the ferroelectric hysteresis loop was measured using an instrument referred as ZT-IA ferroelectric measurement system. Magnetization measurement was carried out on a Quantum Design physical property measurement system (PPMS-9). The frequency dependent dielectric spectrum was measured upon the pellet with electrodes using a IM 3533-01 LCR instrument. The magnetocapacitance parameters of the polycrystalline $La_{0.2}Pb_{0.7}Fe_{12}O_{19}$ ceramic pellets were measured using a IM 3533-01 LCR analyser instrument by applying a variable magnetic field on the specimen.

## 3. Results and Discussion

### 3.1 Structure of $La_{0.2}Sr_{0.7}Fe_{12}O_{19}$

The structure of the as-prepared $La_{0.2}Sr_{0.7}Fe_{12}O_{19}$ powders was detected by an X-ray diffractometer. The heat treatment history of the detecting powders was the same as that of pellet ceramic counterparts with $O_2$ annealing process for three times. Figure 1(b)&(c) show the XRD patterns of the as-sintered $La_{0.2}Sr_{0.7}Fe_{12}O_{19}$ powders and that with subsequent $O_2$ annealing process; the underneath red lines Figure 1(a) are corresponding to the standard diffraction spectrum of pure $SrFe_{12}O_{19}$. It may be seen that all the diffraction peaks of both $La_{0.2}Sr_{0.7}Fe_{12}O_{19}$ patterns match well with the standard red lines of pure $SrFe_{12}O_{19}$, there is no second ferrite phase or any other oxide impurities existing in the spectrum. The charge has already been balanced by substituting 0.3 $Sr^{2+}$ ions with 0.2 $La^{3+}$ ions. The lattice parameters of $La_{0.2}Sr_{0.7}Fe_{12}O_{19}$ are calculated to be a=5.8724 Å and c=23.014 Å in the hexagonal structure with space group *P63/mmc*. The unit cell volume of $La_{0.2}Sr_{0.7}Fe_{12}O_{19}$, however, has been contracted by 8.76 Å$^3$ (0.59%) in comparison with that of $SrFe_{12}O_{19}$. The overlap of the two diffraction patterns suggests that $La_{0.2}Sr_{0.7}Fe_{12}O_{19}$ shares the same structure with $SrFe_{12}O_{19}$ and 0.2 La atoms have successfully substituted 0.3 Sr atomic sites of the $SrFe_{12}O_{19}$ lattice cells, leaving 10% sites vacant.

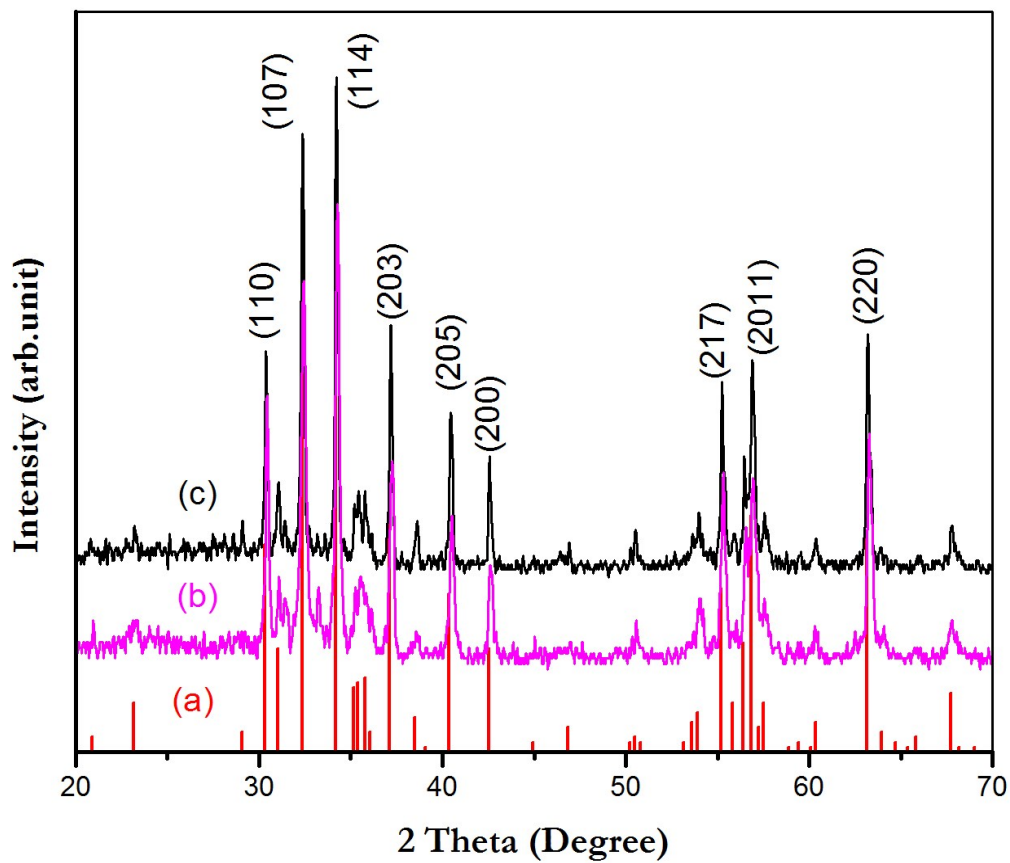

Figure 1: (a) standard diffraction lines from JCPD #33-1340 card, (b) XRD pattern for as-sintered $La_{0.2}Sr_{0.7}Fe_{12}O_{19}$ powders and (c) XRD pattern for $O_2$ annealed $La_{0.2}Sr_{0.7}Fe_{12}O_{19}$ powders.

## 3.2 Antiferroelectric Performance of $La_{0.2}Sr_{0.7}Fe_{12}O_{19}$

Ferroelectricity with large polarization has been observed in $O_2$ annealed $SrFe_{12}O_{19}$ ceramics, whose P-E hysteresis loop takes a classic saturated shape.[31] The phase transition from ferroelectricity to antiferroelectricity takes place at 174°C 错误!未定义书签。. Therefore, the $O_2$ annealed $SrFe_{12}O_{19}$ ceramics demonstrate pure ferroelectricity in form of one full saturated P-E loop at room temperature, which is far below its Curie Temperature (T<<$T_C$). With La substitution for Sr, $La_{0.2}Sr_{0.7}Fe_{12}O_{19}$ ceramic pellet displays double hysteresis loops with antiferroelectric (AFE) characterization, as being shown in Figure 2. The saturated maximum polarization reaches as high as 154 $\mu C/cm^2$, the remnant polarization is around 38 $\mu C/cm^2$, the forward switching (AFE-to-FE) field $E_F$ locates at 25.8 kV/m, backward switching (FE-to-AFE) field $E_A$ is around 35.8 kV/m and switching hysteresis $\Delta E = E_F - E_A$=20 kV/m.

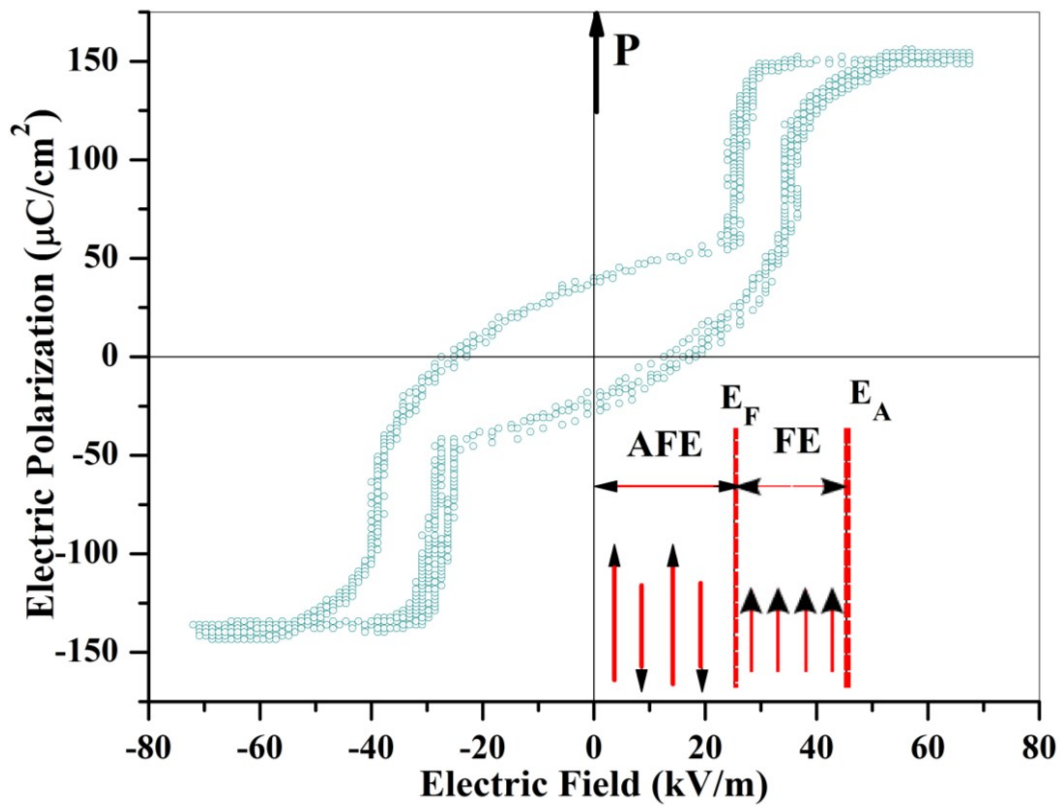

Figure 2: The antiferroelectric polarization hysteresis (P-E) loop of $La_{0.2}Sr_{0.7}Fe_{12}O_{19}$ ceramic, which has been sintered at 1150°C for 1 hour and subsequently annealed at 800°C in pure oxygen for a total duration of 9 hours in 3 steps wise. The measurement was made at a frequency of 33Hz and room temperature (300K).

Both the forward and backward switching fields ($E_A$ & $E_F$) of the $La_{0.2}Sr_{0.7}Fe_{12}O_{19}$ ceramics are much smaller than that of classical antiferroelectric perovskites, such as doped and undoped

Pb(Zr$_x$Ti$_{1-x}$)O$_4$ [40, 41, 42, 43] as well as AgNbO$_3$ [44], whose E$_A$ ranges from 40 kV/cm to 170 kV/cm and E$_F$ from 100 kV/cm to 230 kV/cm. The recoverable energy density, which is expressed as

$$w_{rec} = \int_{P_r}^{P_{max}} EdP$$

where $E$ is the applied electric field, $P_{max}$ is the maximum polarization, and $Pr$ is the remnant polarization, is calculated to be 0.218 J/cm$^3$ for the La$_{0.2}$Sr$_{0.7}$Fe$_{12}$O$_{19}$ ceramic capacitor with electrodes on the opposite surfaces. This value is around 1 order lower than that of perovskite antiferroelectric counterparts. [43, 44] Although this compound is not a good material for energy storage, it is, however, a good candidate for multiferroic memories or logical operation units, since a very low voltage could trig large polarization and thus the energy consumption of the electric devices could be largely reduced.

The middle polarization part between the field range of –E$_F$ to +E$_F$, however, doesn't exhibit clean antiferroelectric feature and still contains a small loop contributing from ferroelectricity. Within the pure antiferroelectric phase or domains, the positive polarization order vectors have the same amplitude as the negative ones but are opposite in direction, the macroscopic P exhibits a linear function relationship with E. The case seems to be different in the antiferroelectric part of Figure 2, where the positive polarization orders show larger amplitude than negative ones and thus induce a mixture feature of antiferroelectric (AFE) and ferroelectric (FE) contributions.

The substitution of La$^{3+}$ for Sr$^{2+}$ in SrFe$_{12}$O$_{19}$ could change the shape of the hysteresis loops. 0.2 La$^{3+}$ replacing 0.3 Sr$^{2+}$ in the unit cell could keep the charge balanced, but induces 10% vacancies in each lattice cell. This leads to the  disruption of translational periodicity of the lattice. La$^{3+}$ substitution and the associated 10% lattice vacancies are known to break the long-range interactions between ferroelectric clusters, as it did in (Pb,La)(Zr,Ti)O$_3$. [45] Above a critical amount of La$^{3+}$ substitution, decoupling is sufficiently strong that the long-range ferroelectric clusters are destroyed and nano-polar domain clusters are established. [46] The doping effect thus turns the properties of SrFe$_{12}$O$_{19}$ from normal ferroelectric to antiferroelectric phase, since substitution of La$^{3+}$ for Sr$^{2+}$ in SrFe$_{12}$O$_{19}$ induces the distortion of the unit cells for easy of domain switching and decrease the oxygen vacancies, which is similar to La$^{3+}$ doped PZT. [47] La$^{3+}$ is also known to be able to stabilize the AFE phase near the AFE-FE morphotropic boundary. [48, 41] Therefore, La$^{3+}$ doped SrFe$_{12}$O$_{19}$ (La$_{0.2}$Sr$_{0.7}$Fe$_{12}$O$_{19}$) shows slim and slanted double hysteresis loops on a macroscopic scale, behaving as a characterization of antiferroelectric phase. These experimental results for the antiferroelectricity of La doped SrFe$_{12}$O$_{19}$ (La$_{0.2}$Sr$_{0.7}$Fe$_{12}$O$_{19}$) are in good agreement with the first principle calculation result, which predicts the frustrated antiferroelectricity of M type hexaferrite.[49] The theory associated the AFE feature of $BaFe_{12}O_{19}$ with its trigonal bipyramidal $Fe^{3+}$ sites and could be made stable at room temperature by appropriate element substitution or strain engineering[49], which matches the case of our La substituted $SrFe_{12}O_{19}$.

### 3.3 Magnetic Properties of $La_{0.2}Sr_{0.7}Fe_{12}O_{19}$

The magnetic property of $La_{0.2}Sr_{0.7}Fe_{12}O_{19}$ was measured by a Physical Property Measurement System (PPMS) at room temperature. The specimen was sintered at 1100°C for 1h and annealed at 800°C in pure $O_2$ atmosphere for 9 hours. Figure 3 displays the ferromagnetic hysteresis M-H loop of $O_2$ annealed $La_{0.2}Sr_{0.7}Fe_{12}O_{19}$ (b) together with that of pure $SrFe_{12}O_{19}$ compounds (a). The coercive field ($H_C$) of $La_{0.2}Sr_{0.7}Fe_{12}O_{19}$ compound is determined to be 5876 Oe, while its remnant magnetic moment (M) is estimated to be 52 emu/g. The corresponding $H_C$ and M values of $SrFe_{12}O_{19}$ compound were measured to be 6070 Oe and 30.7 emu/g, respectively.[31] Both compounds experienced the same heat treatment history. The coercive field ($H_C$) is almost the same for both compounds, but the remnant magnetic moment (M) of $La_{0.2}Sr_{0.7}Fe_{12}O_{19}$ has been enhanced more than 69% in comparison with that of pure $SrFe_{12}O_{19}$ compound, since the substituting $La^{3+}$ provides one more unpaired electron spin. Such promotion of magnetic polarization by La substitution was also observed in La doped $PbFe_{12}O_{19}$ ($La_{0.2}Pb_{0.7}Fe_{12}O_{19}$) after annealing in $O_2$ atmosphere[32, 30].

A strong magnetic feature of $La_{0.2}Sr_{0.7}Fe_{12}O_{19}$ can be assessed by its larger hysteresis loop than that of $SrFe_{12}O_{19}$.[31] With both great ferromagnetic properties and excellent electric properties, $La_{0.2}Sr_{0.7}Fe_{12}O_{19}$ is a kind of typical multifunctional material which simultaneously integrates several functions of antiferroelectrics, ferromagnetism and electro-optics into one structure. The combination of multiple functions in one phase could be able to trig the generation of new chips for electric devices, which would be able to realize multiple functions, such as information storage, logical operation, calculation, optical sensor and magnetoelectric actuation, on only one unit.

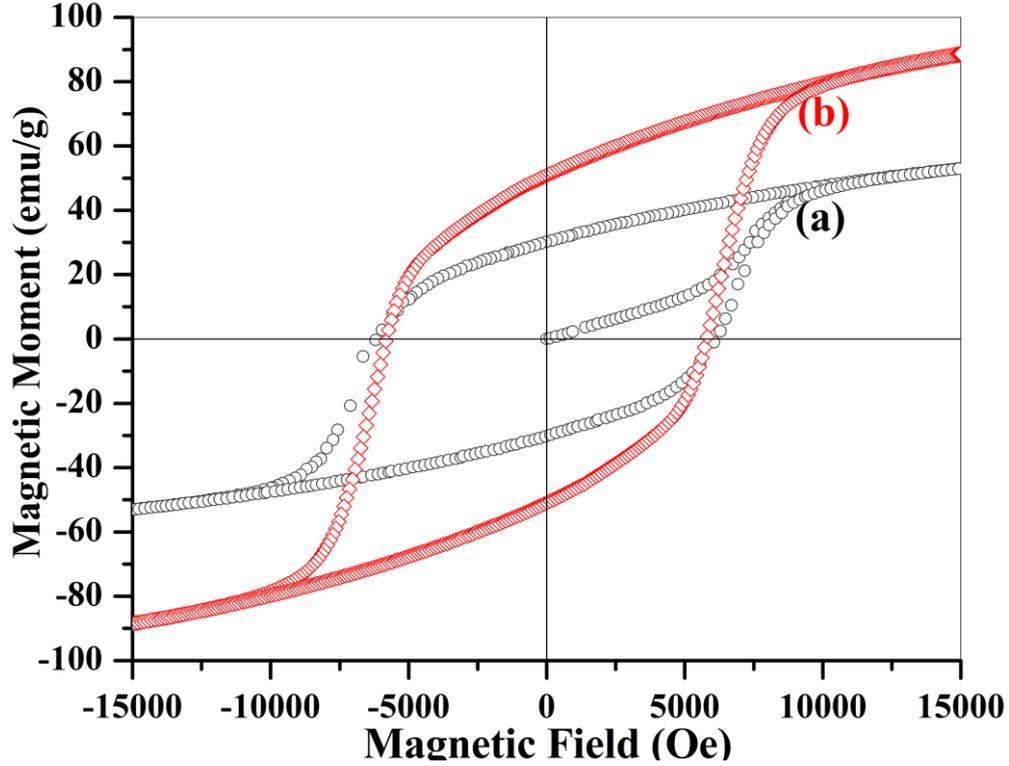

Figure 3: Magnetic hysteresis loops of (a) $SrFe_{12}O_{19}$ and (b) $La_{0.2}Sr_{0.7}Fe_{12}O_{19}$ compounds, which were sintered at 1150°C for 1 h and subsequently annealed in $O_2$ for 9 hours. The measurement was carried out at room temperature.

### 3.4 Magnetic Dielectric Response of $La_{0.2}Sr_{0.7}Fe_{12}O_{19}$

Previous studies on certain rare-earth manganites[9, 50] [6, 42], Y-type[22] and M-type hexaferrites[35, 31] suggested that materials having long wavelength magnetic structures often exhibit a strong interplay between magnetic ordering and ferroelectricity, which makes the capacitance of the manganites[17] and Y-type hexaferrites exhibit great response to the B field[19]. In our previous work, giant magnetodielectric (MD) response was observed in $La_{0.2}Pb_{0.7}Fe_{12}O_{19}$ ceramics, whose magnetocapacitance oscillates at an amplitude of over $1.9 \times 10^5\%$[32]. Such MD response also appears in pure $SrFe_{12}O_{19}$ with much smaller oscillation amplitude[31]. In order to check out if La modified M-type strontium hexaferrite ($La_{0.2}Sr_{0.7}Fe_{12}O_{19}$) could also generate such magnetoelectric coupling response upon an external magnetic field, we again measured the magnetocapacitance as a function of frequency. The $La_{0.2}Sr_{0.7}Fe_{12}O_{19}$ ceramic was coated with silver electrodes on both sides and then placed within a gap between two electromagnets. The electrodes on both surfaces of the ceramic was linked to an IM 3533-01 LCR precision impedance analyser, which would then output the variable capacitance with frequency upon different magnetic field B.

Figure 4 demonstrates such MD response at difference frequencies. Figure 4a&b are the frequency dependence of the dielectric constants $\varepsilon'$ and $\varepsilon''$ upon different magnetic field. Results show that applying magnetic field leads to lift of whole $\varepsilon'$-f spectra and right shif of $\varepsilon''$-f spectra, being similar to the results of $BaFe_{10.2}Sc_{1.8}O_{19}$.[35] Both real and imaginary parts of the dielectric constants vary with frequency as a normal ferroelectric medium. These f-dependent dielectric response matches the Debye relaxation model, which is expressed as follows:

$$\varepsilon'(\omega) = \varepsilon_\infty + (\varepsilon_s - \varepsilon_\infty) \cdot \frac{1}{1+\omega^2\tau^2} \qquad (1)$$

$$\varepsilon''(\omega) = (\varepsilon_s - \varepsilon_\infty) \cdot \frac{\omega\tau}{1+\omega^2\tau^2} \qquad (2)$$

Where $\varepsilon'(\omega)$ and $\varepsilon''(\omega)$ are the real and imaginary parts of dielectric constants, $\omega$ and $\tau$ are frequency and relaxation time, $\varepsilon_s$ and $\varepsilon_\infty$ are dielectric constants at $\omega=0$ and $\omega=\infty$, respectively. At one important frequency point of $\varpi\tau=1$, $\varepsilon'(\omega)=(\varepsilon_s+\varepsilon_\infty)/2$, which is the middle value of the dielectric constant. At this frequency, the imaginary part ($\varepsilon''$) reaches the maximum value, which is $\varepsilon''_m=(\varepsilon_s-\varepsilon_\infty)/2$ and $\omega_m=1/\tau$. The $\varepsilon''$-f spectra in Figure 4b shows a peak with a maximum $\varepsilon''$ at $\omega_m$, which fits this model perfectly. The interesting thing is that applying a magnetic field B lifts the whole $\varepsilon'$-f spectra up and right shifts $\omega_m$ to higher frequency side. The larger is the B field, the bigger is the lift value of $\varepsilon'$ and shift value of $\omega_m$. When $\omega_m = \frac{1}{\tau}\sqrt{\frac{\varepsilon_s}{\varepsilon_\infty}}$, the dielectric loss $\tan\delta_m$ reaches the maximum value. Although magnetic field would lift the whole $\varepsilon'$-f spectra, the increment amplitude of $\varepsilon_s$ at low frequency side is much larger than that of $\varepsilon_\infty$ at high frequency side, as being show in Figure 4 c&d. Therefore, upon application of a B field, the ratio of $\varepsilon_s(B)/\varepsilon_\infty(B)$ would be larger than that of $\varepsilon_s(B_0)/\varepsilon_\infty(B_0)$ [B>B_0>0], as such $\omega_m$ could increases with B and the loss peak would right shift to higher frequency side. The shift values of $\omega_m$ with B are listed in Table 1.

Table 1: Right shift value of the peak position of $\omega_m$ with B field strength.

| Magnetic field B | 0mT | 200mT | 500mT | 800mT | 926mT |
|---|---|---|---|---|---|
| $\omega_m$ for $\varepsilon''$-f spectra peak | 870.9 | 933.3 | 977.2 | 1114.3 | 1513.6 |

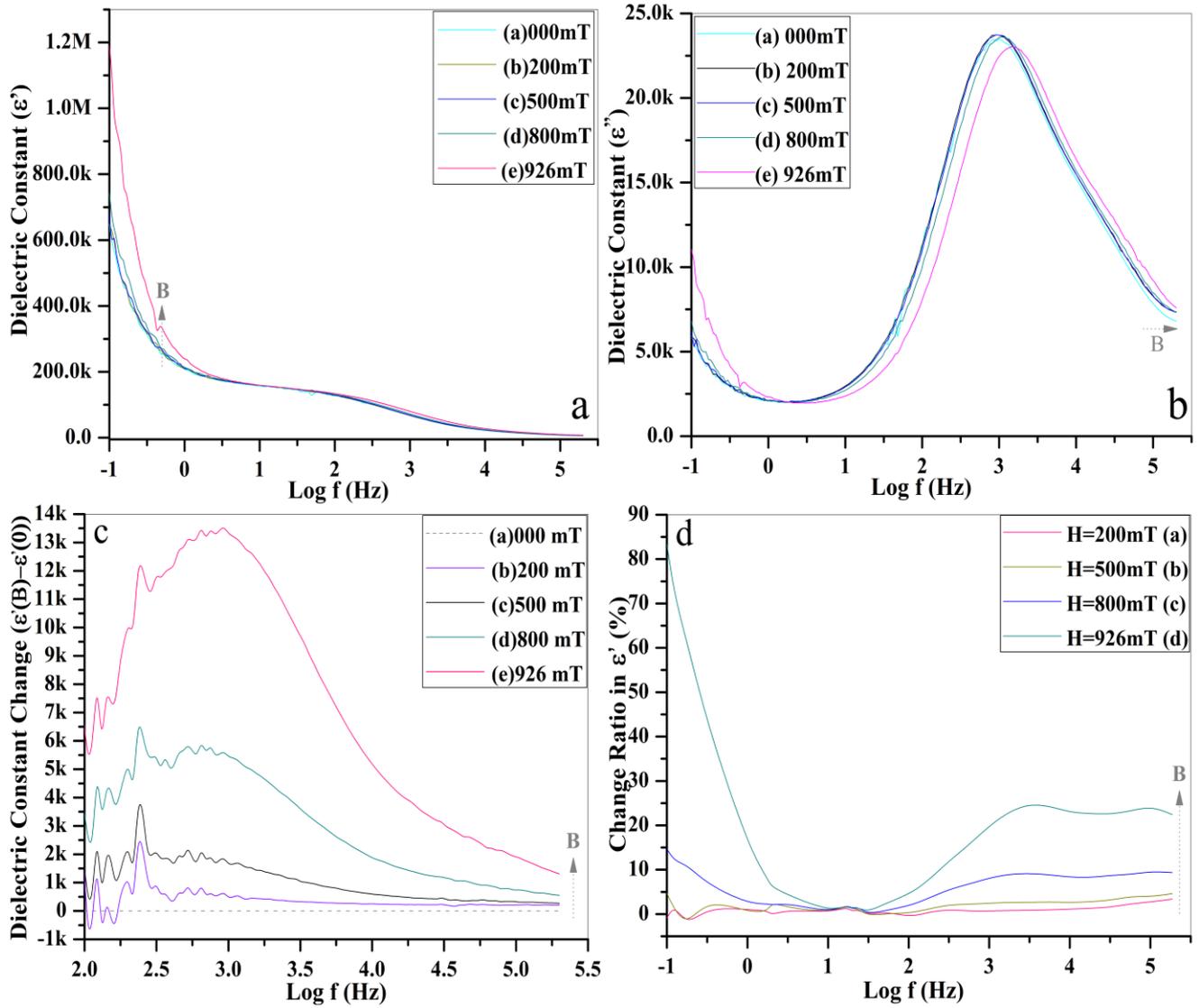

Figure 4: Magnetodielectric response of $La_{0.2}Sr_{0.7}Fe_{12}O_{19}$ ceramics, which displays variable dielectric constants as a function of frequency upon different magnetic fields (B): (a) real part ($\varepsilon^{'}$), (b) imaginary part ($\varepsilon^{''}$), (c) the increment value of dielectric constants ($\varepsilon^{'}$) as a function of frequency upon magnetic field B and (d) MD increment ratio upon B within the whole frequency range (0.1Hz~200kHz).

The variation of dielectric constant depends on B field and frequency which could help us to divide the spectra into four sections (Figure 4d). At first low frequency section (f<10.5Hz), the dielectric constant shows a linear decrement with frequency. The linear slope depends on the magnitude of B field, the higher is B, the larger is the slope. At a fixed frequency, the dielectric constant displays a great enhancement with B field. The whole $\varepsilon^{'}$-f spectra is raised up by applying B field and the change ratio of [($\varepsilon$(B)-$\varepsilon$(0))/$\varepsilon$(0)] increases with B within the first low frequency section, as being shown in Figure 4d. The lower is frequency, the larger is the change ratio. The dielectric change ratio in ε(B) displays a strong MD interplay effect at the first low frequency section. Within the second section between 10.5Hz and 30.5Hz, the dielectric constant is inert to B field and frequency. The dielectric constant displays a slow ascend within the third section between 30.5 Hz and 3.2kHz. Within this section, the absolute MD increment demonstrates a wide peak locating at 3.0kHz and the MD increment value also increases with B field (Figure 4c). At the fourth section of high frequency side (f>3.2 kHz), the variation ratio almost shows a flat line and is independent to frequency for a fixed B field. The change ratio spectrum, however, exhibits great promotion with B field within the whole high frequency section.

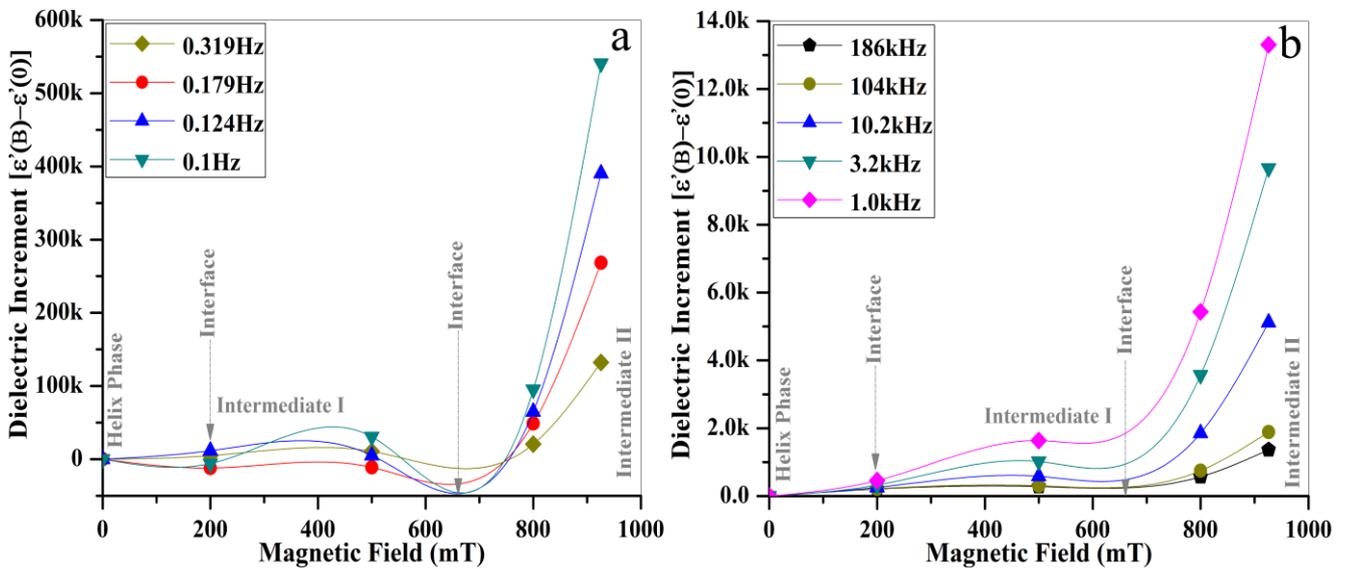

Figure 5: The dielectric constant increment upon magnetic field at different frequencies: (a) low frequency region and (b) high frequency region.

Figure 5 shows the dielectric constant increment upon applied magnetic field at low (a) and high frequency (b) sides, respectively. The MD interplay effect is strong at low frequencies but becomes much weaker at high frequencies. The MD increment reaches as high as 540830 at frequency of 0.1Hz, this value decreases to 13305 at 1kHz and further reduces to 1361 at 186 KHz upon a B field of 926mT. The MD variation ratio is 83.4% at 0.1Hz upon B field of 926mT. This ratio drops to 0.437% at 10.2Hz and again rise up to around 25% within the high frequency range of 3.2kHz~200kHz. The giant oscillation of MD response indicates a strong magnetoelectric coupling effect. The dielectric constant increment keeps almost unchanged within the B field of 0-200mT, then slow ascend appears within the B field range of 200 to

660mT and finally the MD increment displays a sharp ascend within 660 mT to 926 mT. Upon these results, we may be able to configure the magnetic phase structure of $La_{0.2}Sr_{0.7}Fe_{12}O_{19}$, which could be divided into three segments for three successive magnetic phases: helix, intermediate I and intermediate II structures. The two neighboring magnetic phases are separated by an interface, which is marked by a dot grey line. Figure 5 shows the proposed models for the evolution of the magnetic structures by application of B, which is different from that of Y-type hexaferrites[22] and La modified $PbFe_{12}O_{19}$[32]. The first magnetic structure within the B range of 0-200mT is designed to modified helix phase, in which spin vector of $(\boldsymbol{S}_i \times \boldsymbol{S}_j)$ is parallel to $k_0 (e_{ij})$, and thus $P \propto k_0 \times (\boldsymbol{S}_i \times \boldsymbol{S}_j) = 0$. At helix structure, no spin current could be induced by B and therefore the dielectric increment approaches to zero. The intermediate I and II magnetic phases are composed of mainly cycloid conical spins, in which the vector of $(\boldsymbol{S}_i \times \boldsymbol{S}_j)$ is no longer parallel to $k_0 (e_{ij})$, thus additional P would be induced by B[22, 32]. Within the second intermediate I phase (200B<B<660mT), the dielectric increment shows a slow enhancement with B, since B induced small P through cycloid transverse conical spin. This P would produce additional small spin current, which could then enhance the dielectric constants at a small scale. The intermediate II phase (660mT<B<926Mt) could be indicated by a sharp ascend of dielectric constant increment with B, which would induce large P and spin current by cycloid conical spin. These results suggest that $La_{0.2}Sr_{0.7}Fe_{12}O_{19}$ exhibit a strong interplay between magnetic ordering and ferroelectricity, making it a good magnetoelectric coupling candidate.

When a magnetic field B is applied, B would induce a magnetic P(H) through cycloid conical spin in the intermediate phases. Additional P(H) would superimpose on the conventional electric P(E). Therefore, the total polarization P could be expressed as following equation:

$$P_i(E,H) = P_i^s + \varepsilon_0 \varepsilon_r E + \alpha_{ij} H_j + \frac{1}{2} \beta_{ijk} H_j H_k + \gamma_{ijk} H_i E_j \cdots \quad (3)$$

Where $P_i^s$ is the spontaneous polarization, $\varepsilon_0 \varepsilon_{ij} E$ is the conventional electric induced polarization, $\alpha_{ij} H_i$ and $\frac{1}{2} \beta_{ijk} H_j H_k$ is the first order and second order magnetic induced polarization, α and β are first order and second order magnetoelectric coupling coefficient. In case no magnetic field is applied, $P_i(E) = P_i^s + \varepsilon_0 \varepsilon_r E$, which is corresponding to the ε'-f spectra at B=0. When a magnetic field B is applied, additional B induced polarization $P_i(H) = \alpha_{ij}H_j + \frac{1}{2}\beta_{ijk}H_jH_k$ would be added onto P$_i$(E).

For a model of flat capacitor, the capacitance could expressed as following equation:

$$C = \frac{\sigma S}{U} = \frac{\varepsilon_0 \varepsilon_r ES}{U} = \frac{PS}{U} \qquad (4)$$

Where C is the capacitance, $\sigma$ is the polarization charge density, S is surface area and U is the voltage applied on flat capacitor. The dielectric constant is a function of capacitance:

$$\varepsilon_r = \frac{C}{C_0} = \frac{t \cdot C}{\varepsilon_0 S} = \frac{0.036\pi t C}{S} \qquad (5)$$

Where $\varepsilon_r$ is the real part of the dielectric constant, C is the capacitance of the capacitor or the measured ceramic sample, C$_0$ is the capacitance of vacuum ($\varepsilon_0$=8.854×10$^{-12}$$(F/m)$), t is the distance between two electrodes or the thickness of the ceramic sample.

When no B field is applied (B=0), $\sigma = \varepsilon_0 \varepsilon_r E = P(E)$, thus

$$C(E) = \frac{\sigma S}{U} = \frac{P(E)S}{U} \qquad \text{and} \qquad \varepsilon_r = \frac{0.036\pi t P(E)}{U} \qquad (6)$$

Where P(E) is the pure electric induced polarization. This capacitance is corresponding to the $\varepsilon$'-f spectra of B=0. While a magnetic field B is applied, the capacitance is expressed as following equation

$$C(E,H) = \frac{\sigma S}{U} = \frac{P(E,H)S}{U} = \frac{[P(E) + P(H)]S}{U} \qquad (7)$$

$$\varepsilon_r(E,H) = \frac{0.036\pi t P(E,H)}{U} = \frac{0.036\pi t [P(E) + P(H)]}{U} \qquad (8)$$

This capacitance is corresponding to the magnetodielectric (MD) $\varepsilon$-f spectra at $B \neq 0$, as being shown in Figure 4 and Figure 5. Upon dielectric measurement by LCR analysing meter, the voltage U keeps invariable when B varies from 0mT to 926mT. Therefore C(E,H) and $\varepsilon_r$(E,H) depends only on P(E,H) by ratiocination from equations 7&8. Obviously, the capacitance or dielectric constant of the La$_{0.2}$Sr$_{0.7}$Fe$_{12}$O$_{19}$ ceramics could be much more enhanced by applying a magnetic field B, since there is an additional P(H) item in C(E,H) expression of equation 8 in comparison with the C(E) expression of equation 6 without P(H) item in case B=0. The increment value of the capacitance from the contribution of P(H) depends on the magnitude of coupling coefficient ($\alpha$&$\beta$) and the strength of applying B field; the larger is the coupling coefficient, the bigger is C or $\varepsilon_r$ enhancement from P(H). Usually, the intermediate I and II structures with cycloid conic spin have big coupling coefficient ($\alpha$&$\beta$), which could induce large P(H) and thus lead to strong magnetodielectric response in $La_{0.2}Sr_{0.7}Fe_{12}O_{19}$ ceramics.

## 4. Conclusion

In summary, by substitution of 0.3 $Sr^{2+}$ ions with 0.2 $La^{3+}$ in $SrFe_{12}O_{19}$, we fabricated a new M-type hexaferrite - $La_{0.2}Sr_{0.7}Fe_{12}O_{19}$ by a polymer precursor method. The substitution has already balanced charge difference, leaving 10% vacancies in one lattice cell. Both the La substitution and vacancies would break the long-range interactions between ferroelectric clusters and thus turns the property of $La_{0.2}Sr_{0.7}Fe_{12}O_{19}$ from normal ferroelectric to antiferroelectric phase, which displays double hysteresis loops. The maximum saturation polarization and remnant one of the antiferroelectric hysteresis loop are 154 $\mu C/cm^2$ and 38 $\mu C/cm^2$, respectively. Large magnetic hysteresis loop was also observed in $La_{0.2}Sr_{0.7}Fe_{12}O_{19}$ due to its strong ferromagnetism. The coercive field ($H_C$) and remnant magnetic moment (M) of $La_{0.2}Sr_{0.7}Fe_{12}O_{19}$ are measured to be 5876Oe and 52emu/g, respectively. The remnant magnetic moment (M) of $La_{0.2}Sr_{0.7}Fe_{12}O_{19}$ has been improved by 45.3% in comparison with that of pure $SrFe_{12}O_{19}$ due to one more unpaired electronic spin from $La^{3+}$. The combination of both antiferroelectric and ferromagnetic orders in one phase indicates the multiferroic feature of $La_{0.2}Sr_{0.7}Fe_{12}O_{19}$. This compound also demonstrates large magnetodielectric response, the capacitance of the $La_{0.2}Sr_{0.7}Fe_{12}O_{19}$ ceramics has been greatly enhanced by applying a magnetic field. The increment of the real part of dielectric constant depends on the frequency and applying B field, the maximum MD variation ratio in $\varepsilon'$ is 83.4% upon a B field of 926mT. The imaginary part changes very little in magnitude, but the $\varepsilon''$-f spectra displays a right shift to higher frequency side with B field. The physical mechanism for the lift of $\varepsilon'$-f spectra and right shift of $\varepsilon''$-f spectra with B field has been discussed in detail from the point of view of Debye relaxation model and magnetoelectric coupling model. The coexistence of antiferroelectricity and ferromagnetism, in combination with the strong magnetodielectric response in one structure makes $La_{0.2}Sr_{0.7}Fe_{12}O_{19}$ become a new member of multiferroic compounds and imply its great potential application in new generation of integrated electric devices.

**Acknowledgement:**

The authors acknowledge the financial support from National Natural Science Foundation of China under contract No. 11774276.

**Author Contribution**: G. T. configured the concepts and wrote the manuscript; Y. H. prepared the samples and made ferroelectric and magnetic measurements.

**Competing financial interests:** The authors declare no competing financial interests.

**Materials & Correspondence:** The correspondence and material requests should be addressed to G. T. (gltan@whut.edu.cn).